\documentclass[aip,jap,reprint,groupedaddress]{revtex4-1}

\usepackage{graphicx}

\usepackage{amssymb}
\usepackage{amsmath}
\usepackage{amsthm}
\usepackage{color}
\usepackage{units}
\usepackage{float}

\providecommand{\E}[1]{\ensuremath{\times 10^{#1}}}

\newcommand{\diff}{\; \mathrm{d}}
\newcommand{\MG}{Mie-Gr\"{u}neisen }
\newcommand{\Gruneisen}{Gr\"{u}neisen }
\newcommand{\Doring}{D\"{o}ring }
\newcommand{\etal}{et al.\@}

\DeclareMathOperator{\REF}{REF}
\newcommand{\CJ}{{\mathchoice{}{}{\scriptscriptstyle}{} CJ}}

\begin{document}

\title{
A complete equation of state for non-ideal condensed phase explosives
}
\author{S.D. Wilkinson}
\email[]{sw561@cam.ac.uk}
\author{M. Braithwaite}
\author{N. Nikiforakis}
\author{L. Michael}
\affiliation{Cavendish Laboratory, University of Cambridge, Cambridge, UK}

\date{\today}

\begin{abstract}
The objective of this work is to improve the robustness and accuracy of
numerical simulations of both ideal and non-ideal explosives by introducing
temperature dependence in mechanical equations of state for reactants and products.

To this end, we modify existing mechanical equations of state to appropriately
approximate the temperature in the reaction zone. Mechanical equations of state
of \MG form are developed with extensions, which allow the temperature to be
evaluated appropriately, and the temperature equilibrium condition to be
applied robustly. Furthermore the snow plow model is used to capture the effect
of porosity on the reactants equation of state.

We apply the methodology to predict the velocity of compliantly
confined detonation waves.
Once reaction rates are calibrated for unconfined detonation velocities,
simulations of confined rate sticks and slabs are performed, and the experimental
detonation velocities are matched without further parameter alteration,
demonstrating the predictive capability of our simulations.
We apply the
same methodology to both ideal (PBX9502, a high explosive with principal
ingredient TATB) and non-ideal (EM120D, an ANE or ammonium nitrate based
emulsion) explosives.
\end{abstract}

\maketitle

\section{Introduction}




This work is concerned with the numerical simulation of detonation waves
propagating in confined non-ideal explosives, which exhibit velocities of
detonation (VoD) lower than the ones predicted by the Zeldovich-von
Neumann-\Doring (ZND) theory. Accurate
calculation of the VoD for a broad range of confining materials is important for
industrial applications such as mining, where a priori knowledge of the
performance of the explosive is necessary in order to optimize blasting
operations.

Numerical simulation is useful only if it is genuinely predictive.
In the context of mining this means that once the computational model is
calibrated for a particular explosive using unconfined detonation data, it can
then be used to predict VoD curves for other confiners, without any further
parameter adjustment. The material
properties and the behavior of an explosive are captured in a mathematical
model by means of the equations of state (EoS) for the reactants and the
products and the reaction rate law. Although they are both important in
characterizing an explosive, in this work we focus on improving the EoS models.

Commonly used EoS models are in \MG form (such as the JWL EoS \cite{Menikoff2015,
Sheffield2009} for the detonation products and the shock \MG EoS for the reactants
\cite{Wilkins2013, Sheffield2009}) and they relate pressure, volume and energy. These EoS models
are considered to be incomplete because they do not involve the temperature.
However, there are strong motivations for using a temperature
capable EoS, not least because these are necessary for temperature-dependent
reaction rate laws. Even if a pressure-based rate law is used, we have found
that in comparison to incomplete EoS models these allow for more robust implementations of reduced multi-phase formulations
\cite{Michael2015a, Banks2008a, Stewart2007} which employ a
temperature equilibrium condition between reactants and products which coexist
in the reaction zone. Moreover, to ensure the existence of physical solutions to the
temperature equilibrium equation, we require EoS models with which to recover
temperatures.

Important previous work on this topic includes the paper by Wescott \etal
\cite{Wescott2005} (referred to as the WSD model), which presented temperature
capable EoS models for modeling the reactants and products of PBX9502. Those
formulations consist of a mechanical EoS of \MG form with an additional
temperature reference curve. In particular they use the relationship between
the \Gruneisen gamma and the variation in temperature along the reference curve
of the \MG EoS, a relationship which only holds if the reference curve is
isentropic \cite{Davis1998,Davis2000,Wescott2005}.

An alternative approach for calculating temperatures is presented by
Menikoff \cite{Menikoff2009} using a thermal model derived from a vibrational
spectrum from Raman scattering. Temperatures calculated in this way have also
been leveraged for EoS calibration by Aslam \cite{Aslam2017}.
Kittell and Yarrington \cite{Kittell2016}, on
the other hand, derive an expression for post-shock temperatures using a
multi-term Einstein oscillator model for the specific heat capacity.
In either case, the approach is reliant on data which for many explosives are
unavailable.

Similar to the works referenced above, the objective of the present work is to
improve the robustness and accuracy of numerical simulations by introducing
temperature dependence in mechanical EoS models for reactants and products.
We focus on EoS models applicable to both ideal and non-ideal explosives.
We use the ANFO based emulsion EM120D as an example of a
non-ideal explosive; the VoD in
narrow rate sticks can deviate from the Chapman Jouguet (CJ) velocity by as much as 35\% (compare
with the 5\% deviation observed in PBX9502). Note however that PBX9502 may also
be considered an insensitive high explosive which is nearly ideal, as there are other explosives for which the deviation
from the CJ velocity is even less.
Thus the EoS of the products must be
valid even for states at lower entropy than the CJ or principal isentrope. This is
achieved using the ideal detonation code which allows the \Gruneisen gamma to
be evaluated as a function of volume, which is not easy to do using
experimental techniques.
Furthermore EM120D is
porous (14\% by volume) and as such a porosity model is necessary to accurately
predict post-shock temperatures.

To this end, we introduce a temperature reference curve in the \MG EoS.
An ideal detonation code IDeX \cite{Braithwaite1996} is used to
calculate data for the principal isentrope. As such it is possible to use
temperature data to directly calibrate a temperature reference curve for the principal
isentrope, analogous to the pressure reference curve of JWL. In addition, the
\Gruneisen gamma can be evaluated more accurately and so the EoS is valid for
states farther from the reference curve as compared to the JWL EoS which
assumes a constant value for the \Gruneisen gamma.

With regard to the EoS of the reactants, we present an adaptation of the
approach presented by Davis \cite{Davis2000} to construct an isentropic
reference curve from shock Hugoniot data.
The introduction of an explicit porosity model has as an added benefit that a
simple linear fit (as opposed to the nonlinear fit used by Wescott \etal \cite{Wescott2005})
is sufficient to capture the relationship between shock speed and impact speed.

A method is presented to calibrate a temperature capable EoS
which in contrast to the WSD model does not require explicit shock temperature
data. This relies on the assumption that the specific heat capacity at constant volume is
constant.

For both reactants and products a fitting form for the temperature reference
curve consisting of a power law and an exponential term is used. Note that for
an ideal gas at constant entropy the temperature and pressure can be expressed as a power law in
the volume. So for large volumes, for which the exponential term is negligible,
the power law dominates and the material behaves like an ideal
gas, with
constant adiabatic gamma and \Gruneisen gamma.
This fitting form has the desirable property that the \Gruneisen
gamma is well-behaved for all volumes: it remains positive everywhere and does
not diverge. As a result, the implementation of the thermal equilibrium
condition is much more robust, than when using expressions for the \Gruneisen
gamma which do diverge.


In Section~\ref{sec:eos} we present EoS models suitable for detonation
modeling in which the temperature equilibrium closure law is used.
These are discussed with particular focus on their use
in the context of the MiNi16 formulation \cite{Michael2015a}. In
Section~\ref{sec:mod1D} the use of the EoS models
in the one dimensional ZND model is examined. In Section~\ref{sec:pred_mod} the model
is applied to the simulation of rate sticks and slabs, with the aim of
correctly reproducing experimentally measured detonation velocities. Finally
our conclusions are presented in Section~\ref{sec:conc}.

\section{Equations of State}
\label{sec:eos}


The evolution of the fluid is calculated using forces arising
from gradients in the pressure field, so an equation of state (EoS) relating the
pressure, $p$, to the conserved variables is required, such that
\begin{align}
	p = f(v, e),
\end{align}
where $v$ is the specific volume, and $e$ is the specific internal energy.

A standard form for this purpose is the \MG EoS \cite{Menikoff2016}
given as
\begin{align}
	p - p_{\REF}(v) &= \frac{\Gamma(v)}{v} (e - e_{\REF}(v)). \label{eq:MG}
\end{align}

EoS models of \MG form use reference functions, $p_{\REF}(v)$ and
$e_{\REF}(v)$, of arbitrary form and complexity to encode the behavior of the
material at hand. The reference curves specify a one dimensional path through
the (two dimensional) EoS. States which are off the reference curve are
approximated with what is effectively a first order Taylor expansion of the
state at constant volume, which relates the deviation in pressure from the reference curve
with the deviation in specific internal energy. This is done using the
\Gruneisen gamma, $\Gamma$, defined as
\begin{equation} \Gamma(v) = v \left( \frac{\partial p}{\partial e} \right)_v.
\label{eq:gruneisen} \end{equation}
The \Gruneisen gamma is in general a function of volume and entropy. However, equations
of state of \MG form approximate it as a function of the specific volume
only.
This approximation means that the EoS will only be valid for states that are
relatively close to the reference curve.
The reference curve should therefore be chosen to be a locus of states which are
representative of the expected evolution of the material.

For the reactants, the commonly used shock \MG EoS \cite{Wilkins2013, Sheffield2009} uses the
Hugoniot curve as the reference curve, so that the state will remain on
or close to the reference curve when the material is shocked.
Its form is based on the assumption that the shock propagation velocity, $D$,
is linearly related to the flow velocity behind the shock, $u$, such that
\begin{equation}
	D = a + bu,
	\label{eq:linear_Du}
\end{equation}
where $a$ is the ambient speed of sound, and $b$ is a constant fitted to
experimental data.
This approximation fits experimental data well for most solids.
It cannot
however be applied universally, and can be problematic for example in the presence of multiple
shocks \cite{Menikoff2007}.
In the absence of data, the \Gruneisen gamma, $\Gamma$, is
assumed to satisfy
\begin{equation}
	\rho \Gamma = \rho_0 \Gamma_0,
\end{equation}
where $\rho$ is the density and
$\Gamma_0$ is the value at ambient density, $\rho_0$. This is a crude approximation,
and thus the EoS is of limited use for modeling phenomena
where the state deviates significantly from the reference curve.

The detonation products will expand adiabatically in a rarefaction wave after
the chemical reaction is completed. A typical choice for the reference
curve in the product EoS is therefore the adiabat corresponding
to the rarefaction following an ideal detonation wave - the so-called
`principal' isentrope. Note that the adiabat is not necessarily isentropic,
since the gaseous mixture may continue to react or undergo phase changes
within
the rarefaction. However, by taking the adiabat to be the isentrope of the
detonation products, any further reactions or phase changes are implicitly
accounted for in the product EoS. Cylinder tests can be used to measure the
form of the principal adiabat (henceforth called the principal isentrope) in
pressure-volume space experimentally.

The commonly used EoS for detonation products is the JWL
(Jones-Wilkins-Lee) EoS \cite{Menikoff2015, Sheffield2009}, which fits the principal isentrope
data to curves using a combination of one or two exponential terms and a power
law in the volume.


Since no information is available for states away from the isentrope, a
constant value for the \Gruneisen gamma is assumed. This is not strictly valid,
but it is the best approximation that can be made when using cylinder test data
\cite{Davis2001,Short2010}.
Further problems with the JWL EoS are presented by Braithwaite and Sharpe
\cite{Braithwaite2014}. The results show that the \Gruneisen gamma is dependent
on volume, and the constant gamma approximation used by JWL is invalid. This is
of particular importance for non-ideal detonations for which states will be
farther from the reference curve.

\subsection{Temperature}

As discussed in the introduction, many formulations use a temperature
equilibrium condition to allow for the mixing of materials governed by distinct EoS
models.
If this equilibrium condition is to be implemented,
mechanical EoS models need to be extended so that the temperature of
the materials can be approximated.

One approach to extend mechanical EoS models of \MG form, is to supplement them with a
reference function for
temperature, $T_{\REF}$, analogous to the standard reference functions for
pressure and energy. The temperature reference function is closely linked to
the \Gruneisen gamma since
\cite{Menikoff2016,Davis1998,Davis2000,Wescott2005,Menikoff2007}
\begin{align}
	\Gamma = v \left( \frac{\partial p}{\partial S} \right)_v
		\left( \frac{\partial S}{\partial e} \right)_v
		= -\frac{v}{T} \left( \frac{\partial T}{\partial v} \right)_S,
		\label{eq:Gamma1}
\end{align}
where use has been made of one of Maxwell's relations
\begin{equation}
	\frac{\partial^2e}{\partial S \partial v} =
		- \left( \frac{\partial p}{\partial S} \right)_v =
		\left( \frac{\partial T}{\partial v} \right)_S.
\end{equation}
Note that the partial derivative in \eqref{eq:Gamma1} is a derivative at
constant entropy, $S$. The use of this relationship therefore relies on
the reference curve being an isentrope.

Temperature values for states which deviate from the reference curve can be
approximated by relating the temperature change with the change in specific
internal energy using the specific heat capacity at constant volume, $c_v$,
\begin{equation}
	T - T_{\REF}(v) = \frac{e - e_{\REF}(v)}{c_v}.
\end{equation}


This approach defines an EoS with temperature without having to
explicitly calculate any entropies. The fundamental assumption made here
is that any changes in entropy (at constant volume) can be modeled using
entropy-independent values for the \Gruneisen gamma and specific heat
capacities. Thus, the validity of these approximations relies on the entropy changes
being small.

Note that in the present work the specific heat capacity at constant volume,
$c_v$, is assumed to be constant. This assumption has been shown to be incorrect
in some cases \cite{Braithwaite2014}. An avenue for future
investigation is to evaluate whether a volume-dependent or
temperature-dependent specific heat capacity would lead to a significant
improvement in the capability of the model.

\subsection{Products}
Cylinder test experiments only provide data for the pressure and energy
reference curves of the EoS. Hence, data must be
leveraged from elsewhere if a complimentary reference function for the
temperature is to be constructed. An ideal detonation code (\emph{IDeX}) such
as that presented by Braithwaite \etal \cite{Braithwaite1996} can be used for this
purpose. This code uses fluid EoS based on an intermolecular Buckingham
alpha exponential 6 potential.

The program uses the chemical composition of an explosive to find the
configuration of molecules which minimizes the Helmholtz free
energy for a given temperature and volume. This also requires knowledge of the
energy content of the explosive. If in practice the energy content is not known
accurately, it can be reverse engineered using an experimental value for the
ideal detonation velocity.
Using empirical EoS models for each of the product chemicals and mixture
rules, an EoS for the product mixture can be constructed. However, it is cumbersome to
use such an EoS directly in hydrocodes. So instead the code is used to output
pressure, energy and temperature data for the principal isentrope, which is
used to calibrate numerical expressions for the reference functions.

Reference functions of the following form
\begin{align}
	p(v)|_{S_{CJ}} &= a v^b + c \exp(-dv) \\
	e(v)|_{S_{CJ}} &= -\frac{a}{b+1} v^{b+1} + \frac{c}{d}
	\exp(-dv) - Q \\
	T(v)|_{S_{CJ}} &= a_Tv^{b_T} + c_T \exp(-d_Tv), \label{eq:TSCJ}
\end{align}
where
$S_{CJ}$ represents the entropy of the CJ state which lies on the principal
isentrope are then fit to the data.
The energy reference function is obtained by integrating the
pressure reference function, with integration constant $Q$ representing the
specific energy in the large volume limit. The value of the
constant is arbitrary, but is by convention chosen to be the specific energy
release associated with the conversion of material from reactants to products.
The specific energy of the reactants in the large volume limit is set to zero.
The energy release associated with the reaction from reactants to products is
thus accounted for directly in the EoS.

The fitting process is done by fitting the high volume data to a power
law first, and then adding the exponential term as a correction such as to also fit
the low volume data and to satisfy the CJ criterion. In the
large volume limit, where the exponential term goes to zero, the presence of
the power law means that the isentrope approximately takes on the properties of
an ideal gas and is well behaved even at volumes far larger than the volume
range of the data used for the fitting.

In order to accurately reproduce the CJ pressure and ideal VoD
it is important that the value and the derivative of the pressure
reference function are exactly reproduced at the CJ state
\cite{Fickett2000, Lee2008}. To this end, the
parameters $c$ and $d$ are fixed in terms of $a$ and $b$ using
\begin{align}%
	p_{\CJ} &= a v_{\CJ}^b + c \exp(-dv_{\CJ}) \\
	\left. \frac{\partial p}{\partial v} \right|_{S,v=v_{CJ}} &=
		ab v_{\CJ}^{b-1} - cd \exp(-dv_{\CJ}) \\
	d &= \frac{abv_{\CJ}^{b-1} - \left. \frac{\partial
		p}{\partial v} \right|_{S,v=v_{CJ}}}{p_{\CJ} - av_{\CJ}^b} \\
	c &= \exp(dv_{\CJ}) \left( p_{\CJ} - av_{\CJ}^b \right).
\end{align}

\begin{figure}
	\centering
	\includegraphics{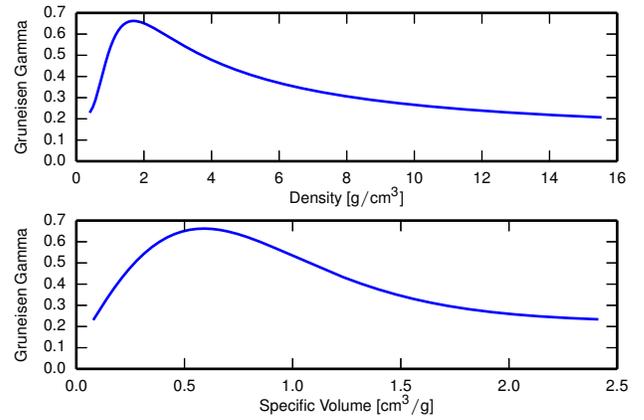}
	\caption{The form chosen for $T_{S_{CJ}}$ ensures that the \Gruneisen gamma
	is bounded and well behaved in both the limit of large volume and the limit
	of large density. The values plotted here are for the EoS of the products
	of PBX9502, the parameters of which are presented in Table
	\ref{tab:products}.}
	\label{fig:gruneisen}
\end{figure}

The form of the function for the \Gruneisen gamma is derived from
\eqref{eq:Gamma1} and \eqref{eq:TSCJ}. The chosen form for $T|_{S_{CJ}}$ is
shown to be suitable in Figure \ref{fig:gruneisen} which demonstrates that for
the products of PBX9502 the \Gruneisen gamma is well behaved across the whole range of
volumes - it never diverges, nor does it go negative or close to zero.
This takes the form:
\begin{align}
	\Gamma(v) = -v \frac{a_Tb_Tv^{b_T-1} - c_Td_T \exp(-d_Tv)}
	{a_Tv^{b_T} + c_T \exp(-d_Tv)}.
	\label{eq:Gruneisen_Tref}
\end{align}
The physical interpretation of the parameters is discussed further in
Section~\ref{sec:reactants} since the same form is used for the reactants.

\begin{figure}
	\centering
	\includegraphics{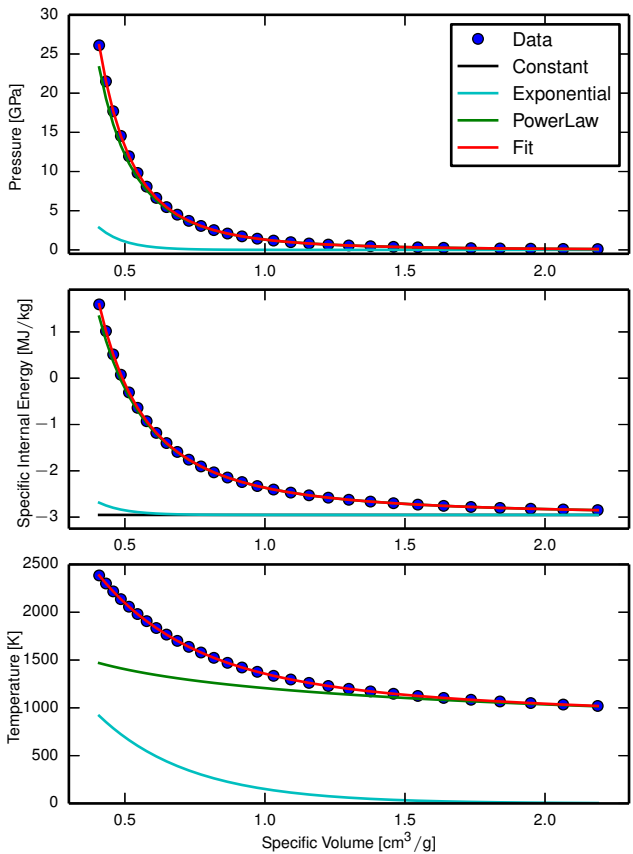}
	\caption{A fit to the principal isentrope from \emph{IDeX} for PBX9502. The
	principal isentrope data are given as blue dots. The
	fit (red) is the sum of the power law (cyan) and the exponential curve (green). The energy
	reference curve also includes a non-zero constant, $Q$, corresponding to the
	energy content of the explosive.
	}
	\label{fig:ref_pbx9502}
\end{figure}

\begin{figure}
	\centering
	\includegraphics{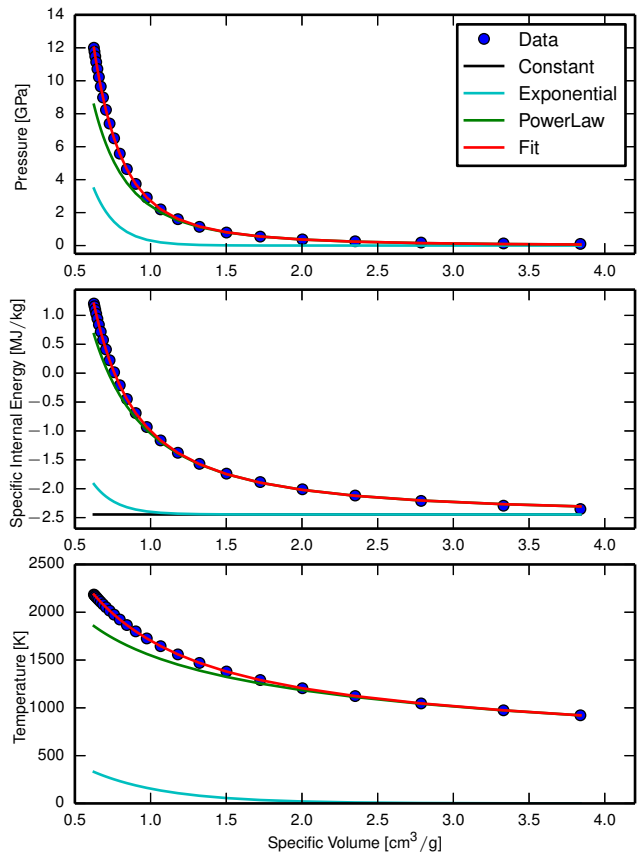}
	\caption{A fit to the principal isentrope from \emph{IDeX} for EM120D.}
	\label{fig:ref_em120d}
\end{figure}

We calibrated EoS models for the products of the two explosives
PBX9502 and EM120D. The parameters are presented in Table \ref{tab:products}.
Note that for PBX9502, the ideal detonation velocity of 7755
$\unit{ms^{-1}}$ from Jackson and Short \cite{Jackson2015} was used to
calibrate the heat of reaction, since \emph{IDeX} predicts a slightly higher
ideal detonation velocity of 7933 $\unit{ms^{-1}}$. For EM120D, on the other
hand, the ideal detonation velocity is taken to be the one which is predicted by
\emph{IDeX}. Figures \ref{fig:ref_pbx9502} and
\ref{fig:ref_em120d} show
the reference curves along with the constituent exponential and power law terms.
This is to show that the presented fitting parameters are such that the power law
is the dominant term, especially in the large volume limit.

\begin{table}
	\centering
	\begin{tabular}{|c|c|c|}
		\hline
		 & PBX9502 & EM120D \\
		\hline
		$a$ & $0.2865$ & $17.47$ \\
		$b$ & $-3.219$ & $-2.712$ \\
		$c$ & $2.233\E{11}\ \unit{Pa}$ & $2.109\E{11}\ \unit{Pa}$\\
		$d$ & $10700\ \unit{kgm^{-3}}$ & $6571\ \unit{kgm^{-3}}$ \\
		$a_T$ & $264.8$ & $107.5$ \\
		$b_T$ & $-0.2195$ & $-0.3861$ \\
		$c_T$ & $3188\ \unit{K}$ & $1171\ \unit{K}$ \\
		$d_T$ & $3053\ \unit{kgm^{-3}}$ & $2025\ \unit{kgm^{-3}}$ \\
		$Q$ & $2.953\ \unit{MJkg^{-1}}$ & $2.446\ \unit{MJkg^{-1}}$ \\
		$c_V$ & $2500\ \unit{JK^{-1}kg^{-1}}$ & $2500\ \unit{JK^{-1}kg^{-1}}$ \\
		$v_{\CJ}$ & $1/2450.2\ \unit{m^3kg^{-1}}$ & $1/1598.4\ \unit{m^3kg^{-1}}$ \\
		$p_{\CJ}$ & $26.12\ \unit{GPa}$ & $12.00\ \unit{GPa}$ \\
		$D_{\CJ}$ & $7755.0\ \unit{ms^{-1}}$ & $6389.5\ \unit{ms^{-1}}$ \\
		\hline
	\end{tabular}
	\caption{Parameters for the product EoS models for PBX9502 and
	EM120D. The proposed reference curves are shown in Figures \ref{fig:ref_pbx9502} and
	\ref{fig:ref_em120d}.
	The first four rows of parameters correspond to the pressure and energy
	reference curves, while the next four (with subscript $T$) correspond to
	the temperature reference curve \eqref{eq:TSCJ}.
	}
	\label{tab:products}
\end{table}

\begin{figure}
	\centering
	\includegraphics{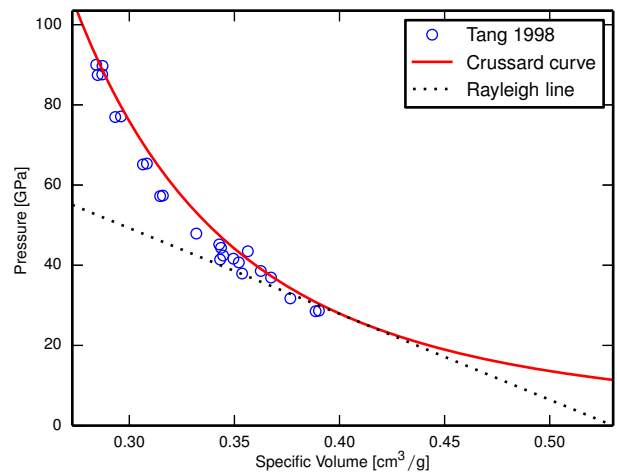}
	\caption{The EoS calibrated using the ideal detonation code data is used to
	calculate the Crussard curve for PBX9502. It matches the experimental data
	from Tang \etal \cite{Tang1998} reasonably well. The CJ state is at the intersection of the
	Crussard curve with the Rayleigh line.}
	\label{fig:pbx9502_crussard}
\end{figure}

Having calibrated the EoS to fit the principal isentrope, it cannot be
assumed that the EoS will accurately reproduce the Hugoniot curve for the
products (also called the Crussard curve) for overdriven detonations. These
states are at higher entropy than the principal isentrope, and thus rely on
accurate values for the \Gruneisen gamma, as well as an accurate isentrope.
Figure \ref{fig:pbx9502_crussard} shows that the EoS for PBX9502 matches the
overdriven detonation data from Tang \etal \cite{Tang1998} reasonably well. This serves to
validate the calibration.

\subsection{Reactants}
\label{sec:reactants}
A methodology presented by Davis \cite{Davis2000} can be used to construct an
EoS for explosive reactants using an isentropic reference curve. An equation
for the isentrope pressure is derived using the assumption of a linear $D$,$u$
relationship \eqref{eq:linear_Du} as is used in the shock \MG EoS
\begin{align}
	D &= a + bu \label{eq:Dulinear} \\
	p_{\REF}(v) &= \frac{\rho_0 a^2}{4b} \left[
		\exp(4b(1-v/v_0)) - 1 \right] \\
	e_{\REF}(v) &= \left(\frac{a}{4b}\right)^2 \left(\exp(4b(1 - v/v_0)) - 1
	\right) \nonumber\\
		& \qquad \qquad \qquad + \frac{\rho_0 a^2}{4b} (v - v_0).
\end{align}
Note that these expressions for the pressure and energy reference curves do not
diverge in the limit of small volumes, but grow sufficiently quickly to avoid
potential practical issues.

Data for the volume and pressure on the Hugoniot curve can be used to
calibrate the \Gruneisen gamma. Across a shock wave the entropy increases,
and thus states on the Hugoniot curve lie above the reference curve which is an
isentrope. The deviation between the Hugoniot curve and the isentrope is
related to the \Gruneisen gamma. Note, however, that in practice this calibration
process is somewhat ill-conditioned. Small relative errors in pressure
measurements for Hugoniot states become more significant when the isentrope
pressure is subtracted from it. The fitting process must therefore be carried
out with care.

The \Gruneisen gamma must not diverge or go negative for a thermodynamically
stable EoS \cite{Segletes1991}.
To ensure that this is the case, we fit it with the same form that was used for
the products EoS. The reference temperature is defined as a power law with a
correcting exponential at small volumes \eqref{eq:TSCJ}. The parameters of
$T_{\REF}$ are restricted to ensure that the EoS behaves
normally: $a_T$, $c_T$ and $d_T$ are set to be positive, while $b_T$ must be
negative. The values of the parameters can be further constrained by observing
that $-b_T$ is the \Gruneisen gamma in the limit of large volumes, where the
EoS begins to behave like an ideal gas. As such we expect $-b_T$
to have a value close to the ambient \Gruneisen gamma, $\Gamma_0$,
\begin{equation}
	\Gamma_0 = \frac{\beta c^2}{c_p},
\end{equation}
where $\beta$ is the ambient coefficient of thermal expansion, $c$ is the
ambient frozen sound speed and $c_p$ is the ambient specific heat capacity at
constant pressure.
Furthermore we choose initial values for the fitting process such that the
power law is the dominant term. As a result, the EoS will approach ideal gas-like
behavior in the large volume limit.

\subsubsection{Porosity Model}

For porous materials of total specific volume, $v_0$, we define the crushing specific
volume, $v_{00}$, to be the specific volume of the matrix material - the material
excluding the pores. Compression of the material at low densities requires
little work, since it principally leads to closing of the pores, while the
density of the matrix material remains largely unchanged.
Compression at higher density, on the other hand, leads
to compression of the matrix material and requires more work.

The leading shock at the front of a detonation wave compresses the material to
volumes significantly smaller than the crushing volume. As such it is adequate
to adapt the reference curves of the reactant EoS following the
snow plow model \cite{Zeldovich1965, Handley2010}.
The compressibility of the material at volumes larger than the crushing
specific volume is taken to be
infinity. In other words, it is assumed that the integral of pressure with
respect to volume, which represents the work done compressing the material, is
entirely due to the work done for compression beyond the crushing volume.
The reference pressure on the isentrope is thus taken to be zero for
larger volumes.

\begin{figure}
	\centering
	\includegraphics{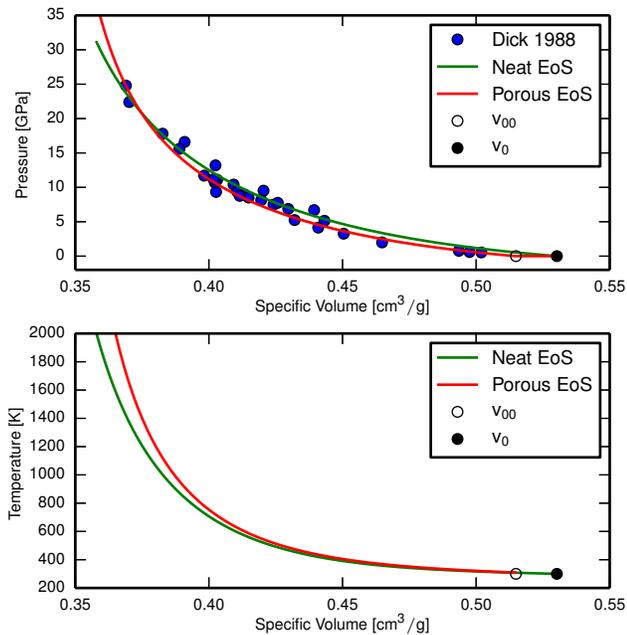}
	\caption{Comparison of Hugoniot curves when using the porosity model, and when
	using the standard EoS.
	The EoS models are compared with experimental data for PBX9502
	\cite{Dick1988, Gustavsen2006}.
	Both EoS models are calibrated using the same volume-pressure
	Hugoniot data. The predicted temperatures increase as a result of using the
	porosity model.
	}
	\label{fig:porous_hugoniots}
\end{figure}

It is possible to calibrate the EoS such as to match the volume-pressure
experimental data for the Hugoniot curve even without employing any porosity
model.
However the temperature and $D$,$u$ relationship is significantly
affected by the explosive porosity. Figure \ref{fig:porous_hugoniots} shows how
the temperature of the Hugoniot path is increased if the porosity of the
reactants is captured using the method presented here. The explosive modeled
is PBX9502 with an initial density of $1886\ \unit{kgm^{-3}}$. The density
corresponding to the crushing specific volume \cite{Dick1988} is taken to be
$1942\ \unit{kgm^{-3}}$.

\begin{figure}
	\centering
	\includegraphics{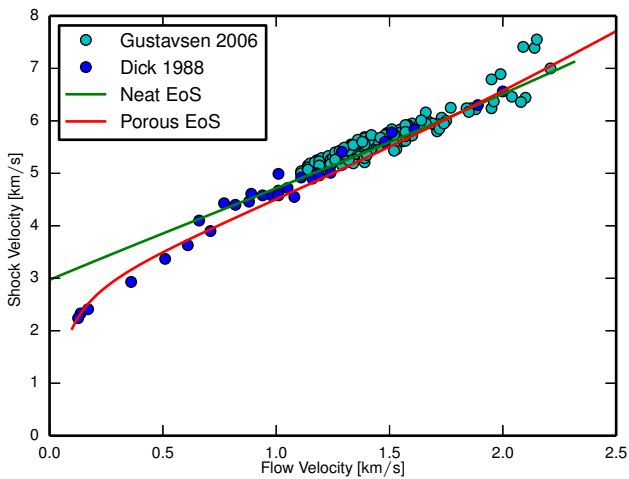}
	\caption{The relationship between shock velocity and flow velocity is
	affected by the porosity model. The linear relationship curves towards zero
	for weak shocks. In the limit of small shocks, this is clearly not valid,
	however for moderate shocks this curve fits the data well
	\cite{Dick1988, Gustavsen2006}.
	}
	\label{fig:porous_hugoniots_du}
\end{figure}

It has been noted before that the $D$,$u$ relationship for PBX9502 is not linear
across all shock velocities \cite{Sheffield2009}.
Figure \ref{fig:porous_hugoniots_du} shows however that an EoS constructed
using a linear fit for the $D$,$u$ relationship and extended with a porosity
model will match the experimental data for moderate as well as strong shocks.
Use of the porosity model here means a simple linear fit of the $D$,$u$ data in
the strong shock regime is sufficient to model a wide range of shock strengths.
The WSD model \cite{Wescott2005}, on the other hand, uses a nonlinear fit to
the $D$,$u$ data.

Modeling porosity in this way can be problematic in the weak shock regime. The
speed of sound in the porous reactants under ambient conditions is unphysical.
This is because the
compressibility of the porous material is of course finite, but we have assumed
it to be infinite. Furthermore the predicted shock velocity is
unphysical for weak shocks. Figure \ref{fig:porous_hugoniots_du} shows that the
Hugoniot curve in the $D$,$u$ space curves towards zero in the limit of small $u$.
As such the snow plow model must be improved upon if for example ignition is to
be modeled.

More complex models of porous materials, such as the $P$-$\alpha$ model capture
the effect of porosity for weak shocks much more accurately \cite{Afanasenkov1969,
Herrmann1969, Carroll1972, Menikoff2000}.

The curved shape of the $D$,$u$ Hugoniot arising from the snow plow model
matches that presented by Lambourn and Handley \cite{Lambourn2017},
Menikoff \cite{Menikoff2007}, and Schoch
\cite[Appendix G]{Schoch2012} where the porosity is modeled explicitly using a
multiphase model. In the multiphase model, the matrix material is represented
by the shock \MG EoS which is calibrated using data for the non-porous
explosive. The pores are modeled using the ideal gas EoS.

\subsubsection{Temperature of Reactants in the Expansion Regime}

The modeling of explosive reactants poses difficulties when states in the
expansion regime occur \cite{Segletes1991}. The data
available for calibration relate exclusively to
states under compression - which is the regime of interest for modeling shock
waves. However states in the expansion regime can occur
in the context of direct numerical simulation of detonation
waves.
The usual location of these states
is far behind the detonation wave where the detonation
products have rarefied and depressurised to ambient conditions. If the loss of
pressure is fast, the explosive may stop burning while a small amount of
reactants is still present. The pressure in these cells must be found by
applying the usual pressure and temperature closure conditions. The EoS models
must therefore be suitable for finding pressure equilibrium and
temperature equilibrium under these conditions, even though the state has much
less energy than typical cells in the reaction zone.

Other authors have also encountered this problem. Arienti \etal \cite{Arienti2004}
developed an approach for dealing with large volume states when using the shock
\MG EoS. Menikoff \cite{Menikoff2009} also introduces a work-around
specifically for the expansion regime. Since states in the expansion regime
will only occur far from the front of the detonation wave, outside the
detonation driving zone, the handling of these states will not influence the
predicted VoD. It is only required
to ensure that the model can be applied robustly across the entire domain of
the simulation.

Given the porosity model discussed above, the pressure reference curve is
chosen to be exactly zero for volumes above $v_{00}$. It is clearly not
isentropic in this regime, as such an additional term (which increases with
volume) must be added to the temperature reference curve for large volumes.
This ensures that the coefficient of thermal expansion,
\begin{equation}
	\beta = \frac{1}{v} \left( \frac{\partial v}{\partial T} \right)_p,
	\label{eq:coeff_thermal_expansion}
\end{equation}
is positive for all volumes, including the expansion regime.
This is essential for robust solution of the thermal equilibrium equations.

The form of this additional term required by the temperature reference curve in
the expansion regime is calculated by considering the difference between the
new $p_{\REF}$ which has been set to zero for the purpose of modeling the
porosity and the original form, $\tilde{p}_{\REF}$,
\begin{equation}
	p_{\REF} - \tilde{p}_{\REF} = \rho\Gamma(v) \frac{T_{\REF} - \tilde{T}_{\REF}}{c_V}.
\end{equation}

The summarized equations for the reactant EoS are
\begin{align}
	\tilde{p}_{\REF}(v) &= \frac{\rho_0 a^2}{4b} \left[ \exp(4b(1-v/v_{00})) - 1 \right] \\
	\tilde{e}_{\REF}(v) &=
		\left(\frac{a}{4b}\right)^2 \left[ \exp(4b(1 - v/v_{00})) - 1
		\right] + \frac{\rho_0 a^2}{4b} (v - v_{00}) \\
	\tilde{T}_{\REF}(v) &= a_Tv^{b_T} + c_T \exp(-d_Tv) \\
	\text{for } v & < v_{00} \left\{ \begin{array}{rl}
		p_{\REF}(v) &\!\!= \tilde{p}_{\REF}(v) \\
		e_{\REF}(v) &\!\!= \tilde{e}_{\REF}(v) \\
		T_{\REF}(v) &\!\!= \tilde{T}_{\REF}(v) \\
		\end{array} \right. \\
	\text{for } v & \geq v_{00} \left\{ \begin{array}{rl}
		p_{\REF}(v) &\!\!= 0 \\
		e_{\REF}(v) &\!\!= 0 \\
		T_{\REF}(v) &\!\!= \tilde{T}_{\REF}(v) - \frac{\tilde{p}_{\REF}}{\rho
			\Gamma(v) c_V} \\
		\end{array} \right.
\end{align}
The equation for the \Gruneisen gamma is the same as for the products EoS
\eqref{eq:Gruneisen_Tref}.

Note that volume-pressure
data for the Hugoniot curve are required to calibrate for the temperature
reference curve of the EoS above. For the emulsion explosive EM120D,
these data are not available. We therefore use the Hugoniot curve as calculated
by Schoch \cite{Schoch2012}, where the shock response of the porous material is
modeled using a multiphase model. The parameters for the reactants of both
explosives are presented in Table \ref{tab:reac}.

\begin{table}
	\centering
	\begin{tabular}{|c|c|c|}
		\hline
		& PBX9502 & EM120D \\
		\hline
		$a$ & $2970\ \unit{ms^{-1}}$ & $2170\ \unit{ms^{-1}}$ \\
		$b$ & 1.81 & 1.82 \\
		$a_T$ & 5.141 & 2.073 \\
		$b_T$ & -0.5371 & -0.6867 \\
		$c_T$ & $258020\ \unit{K}$ & $22805\ \unit{K}$ \\
		$d_T$ & $19960\ \unit{kgm^{-3}}$ & $10660\ \unit{kgm^{-3}}$ \\
		$\rho_{00}$ & $1942\ \unit{kgm^{-3}}$ & $1400\ \unit{kgm^{-3}}$ \\
		$\rho_0$ & $1886\ \unit{kgm^{-3}}$ & $1210\ \unit{kgm^{-3}}$ \\
		$c_V$ & $1000\ \unit{JK^{-1}kg^{-1}}$ & $1000\ \unit{JK^{-1}kg^{-1}}$ \\
		\hline
	\end{tabular}
	\caption{Parameters for the reactant EoS models for PBX9502 and EM120D.}
	\label{tab:reac}
\end{table}

\subsection{Closure Rules for Coexistence of Materials}


Modeling of non-ideal detonation waves requires resolution of the DDZ
(detonation driving zone) --- only cells in this zone play a role in determining
the velocity of detonation. This zone includes part of the reaction zone in
which the chemical reaction occurs.
Since the reactants and products are modeled using independent EoS models, the
coexistence of both materials in the reaction zone requires careful attention.

The mathematical formulation used in this work is the formulation of Michael and
Nikiforakis (MiNi16) \cite{Michael2015a}. In one dimension, the governing
equations for the evolution of the three materials are
\begin{equation}
	\frac{\partial}{\partial t}
	\begin{pmatrix}
		z \rho_1 \\
		(1-z) \rho_2 \\
		\rho u \\
		\rho E \\
		(1-z) \rho_2 \lambda
	\end{pmatrix}
	+ \frac{\partial}{\partial x}
	\begin{pmatrix}
		z \rho_1 u \\
		(1-z) \rho_2 u \\
		\rho u^2 + p \\
		u (\rho E + p) \\
		(1-z) \rho_2 \lambda u
	\end{pmatrix}
	=
	\begin{pmatrix}
		0 \\
		0 \\
		0 \\
		0 \\
		K
	\end{pmatrix}
	\label{eq:governing}
\end{equation}

The three materials correspond to the
confiner (labeled $1$), the reactants ($\alpha$) and the products ($\beta$).
In addition, the reactants and products collectively form the explosive which
is referred to as the second phase (labeled $2$).
Variables without subscript labels refer to the properties of the three
material mixture. There is a single value for the flow velocity, $u$, and the
pressure, $p$. The conversion from reactants to products is encoded in the
reaction rate,
\begin{equation}
	K = \frac{\partial \lambda}{\partial t}.
\end{equation}

The mixing rules for the three materials
are presented here along with
a general EoS of \MG form
\begin{align}
	\rho &= z \rho_1 + (1-z) \rho_2 \label{eq:rho_hybrid_phase} \\
	\frac{1}{\rho_2} &= \frac{\lambda}{\rho_\alpha} +
		\frac{1-\lambda}{\rho_\beta} \label{eq:rho_hybrid_mix} \\
	\rho E &= \frac{1}{2} \rho u^2 + z \rho_1 e_1 + (1-z) \rho_2 e_2
		\label{eq:e_hybrid_mix} \\
	e_2 &= \lambda e_\alpha + (1-\lambda) e_\beta \\
	p_k - p_{k,\REF}(v_k) &= \rho_k \Gamma_k(v_k) (e_k - e_{k,\REF}(v_k))
		\nonumber \\
		& \qquad \qquad \text{for } k \in \{1,\alpha,\beta\}. \label{eq:mie_gruneisen}
\end{align}

Equation \eqref{eq:rho_hybrid_phase} expresses the overall density, $\rho$, in
terms of the phase densities, $\rho_1$ and $\rho_2$, using the volume fraction,
$z$. The density of the second phase is itself a combination of the
densities,
$\rho_\alpha$ and $\rho_\beta$ \eqref{eq:rho_hybrid_mix}. In this case the
relative amount of reactants ($\alpha$) in the explosive mixture (phase 2) is
given by the mass fraction, $\lambda$.
Analogously, the total specific energy, $E$, of the three material system
\eqref{eq:e_hybrid_mix} is defined in terms of the specific internal energies
of the phases, $e_1$ and $e_2$, while the second phase is itself a combination
of the reactant and product specific internal energies, $e_\alpha$ and
$e_\beta$.

A pair of coexisting materials will tend towards pressure
equilibrium and temperature equilibrium. The time scale on which pressure
equilibrium is reached is very fast.
The temperature equilibrium time scale,
however, may be much longer and is not necessarily fast in comparison to the
time scale associated with the detonation wave.
Davis \cite[chap. 3]{Zukas2002} and Stewart \etal \cite{Stewart2002} discuss
the merits of various closure laws that could be used in place of temperature
equilibrium. Matignon \etal \cite{Matignon2012} discuss
various such laws and their effect on detonation shock dynamics.

Following MiNi16 \cite{Michael2015a}, pressure equilibrium is
assumed between phases $1$ and $2$. For the second phase, both pressure and
temperature equilibrium are applied. Thus pressure equilibrium applies to all
three phases, and temperature equilibrium is applied between materials $\alpha$
and $\beta$.

Unfortunately the closure conditions of pressure and temperature equilibrium,
which are introduced to remove the degrees of freedom associated with space and
energy distribution, do not permit a closed form expression for the pressure.
The new equation
\eqref{eq:hybrid_mixed_eos} for the pressure is
derived by substituting the component EoS models into the energy equation
\eqref{eq:e_hybrid_mix}
\begin{widetext}
\begin{equation}
	\rho (E- \frac{1}{2} u^2) = p \left( \frac{z}{\Gamma_1(v_1)} +
		\frac{(1-z) \rho_2 \lambda}{\rho_\alpha \Gamma_\alpha(v_\alpha)} +
		\frac{(1-z) \rho_2(1-\lambda)}{\rho_\beta \Gamma_\beta(v_\beta)} \right)
	+ z \rho_1 \REF_1 + (1-z) \rho_2 \lambda \REF_\alpha +
		(1-z) \rho_2 (1-\lambda) \REF_\beta,
	\label{eq:hybrid_mixed_eos}
\end{equation}
where
\begin{equation*}
	\REF_k = \frac{-p_{k,\REF}(v_k)}{\rho_k \Gamma_k(v_k)} + e_{k,\REF}(v_k)
		\qquad \text{for } k \in \{1,\alpha,\beta\}.
\end{equation*}
\end{widetext}

This equation has two unknowns: the pressure and one or other of $v_\alpha$ and
$v_\beta$ (the other is fully constrained by \eqref{eq:rho_hybrid_mix}).
The pressure equation is solved along with the equation for temperature
equilibrium between the reactants and products
\begin{align}
	T_\alpha &= T_\beta.
		\label{eq:t_eqm}
\end{align}

The solution is found by applying a numerical nonlinear root-finding method.
This is done using a modified version of Brent's method. Brent's method uses the
secant method principally, but resorts to the bisection method under certain
conditions, thus guaranteeing convergence provided a root exists. Here we
replace the secant method with Newton-Raphson. Convergence is therefore
guaranteed, and the rate of convergence will be of second order in most cases.
In practice the great majority of the iterations of the method will proceed
according to the Newton-Raphson method and converge at second order.

If the mixture consists of mainly products ($\lambda < 0.5$) then $v_\alpha$ is
taken as the independent variable, and $v_\beta$ is calculated using
\eqref{eq:rho_hybrid_mix}, while for mixtures with mainly reactants $v_\beta$
is chosen as the independent variable. This is necessary because solving
\eqref{eq:rho_hybrid_mix} for $v_\alpha$ becomes ill-conditioned for very small
$\lambda$. It involves subtracting two numbers of almost equal magnitude.

The set of equations above can be problematic if EoS models of limited validity are
used, and so it cannot be guaranteed in general that a solution will exist.
Furthermore, care should be taken as solutions that do exist may be physically
invalid. In particular
the negative density domain may have mathematical solutions which are
physically meaningless. Depending on the EoS models, regions of negative temperature or
negative pressure can also cause the root-finding algorithm to fail.


Mathematical solutions which represent unphysical, negative density states can
be excluded by restricting the search domain. The restriction can be expressed
as
\begin{equation}
	\rho_\alpha \ge \lambda \rho \quad \mathrm{or} \quad
		\rho_\beta \ge (1-\lambda) \rho. \label{eq:density_restriction}
\end{equation}
In other words the mass of material $\alpha$ in the cell cannot exceed the
total amount of mass in the cell. If $\rho_\alpha$ violates this restriction
the result for $\rho_\beta$ will be negative. This is easily understood when it
is considered that equation \eqref{eq:rho_hybrid_mix} is an addition of
volumes, each of which must necessarily be positive.

This is the fundamental reason that the set of equations is not guaranteed to
have solutions. If equation \eqref{eq:t_eqm} is considered on its own
(assuming the pressure has some fixed value) without any restrictions on the
densities, then it is guaranteed to have solutions for well behaved EoS models.

As the density of a material approaches zero, the temperature will also
approach zero. In doing so the density of the other material will increase, and
its temperature will accordingly increase. At some point these temperature
curves are bound to cross; this is the point of temperature equilibrium.
However this crossing point may be in the region that has been excluded by
the restrictions on density \eqref{eq:density_restriction}.

It is thus not realistic to guarantee that the problem will have solutions for
an arbitrary state. We can however ensure that the equations will have
a solution for all realizable states by ensuring the thermodynamically
consistent behavior of the EoS models in the limits of large and small
volumes.

Firstly, the \Gruneisen gamma must be positive and bounded for all possible volumes
accessed by the simulation. The form of the fitting function chosen for
$T_{\REF}$ ensures that this is the case as shown in Figure
\ref{fig:gruneisen} which demonstrates that the limiting behavior of the
\Gruneisen gamma is appropriate for large volumes as well as large densities.

Secondly, the temperature must increase monotonically with volume at any fixed
pressure. In other words the coefficient of thermal expansion
\eqref{eq:coeff_thermal_expansion} must be positive.
This can be verified for given EoS parameters before running the simulation. For
the product EoS the isobars were found to increase monotonically across all
volumes, while for the
reactants the isobars were found to have a minimum, but at sufficiently low
volumes to play no role in the simulation. To ensure that anomalous roots
at these unphysically small volumes do not appear, smaller volumes are
explicitly excluded from the search domain.




\section{Modeling of Detonation Waves in 1D}
\label{sec:mod1D}
The equations of MiNi16 comprise a system of hyperbolic partial differential
equations \cite{Michael2015a}. The equations are solved numerically using a conservative finite
volume method. By defining the flux with the Godunov scheme \cite{GOD59,TOR09},
the problem is reduced to solving a Riemann problem at each cell interface.
The Harten, Lax and van Leer, Contact (HLLC) approximate Riemann solver is
used. It was first presented by Toro, Spruce and Spears \cite{TOR94} and is an
extension of the HLL method\cite{HAR83}. This is
extended to second order using MUSCL-Hancock with the van Leer limiter
\cite{TOR09}.

The methods described above are implemented in a code developed at the
Laboratory of Scientific Computing at the University of Cambridge. This code is
capable of adaptive mesh refinement (AMR) and parallel execution through
subdivision of the domain. Simulations of rate stick detonations can be greatly
accelerated with adaptive mesh refinement, because the detonation wave has a
complex structure, which is very narrow in comparison to the domain. A high
resolution is required to resolve the detonation wave, but it is impractical to
use the fine resolution for the whole domain.

\begin{figure}
	\centering
	\includegraphics{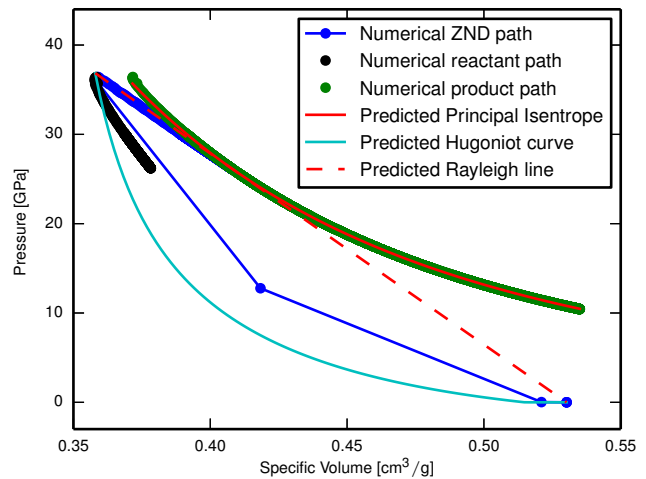}
	\caption{The blue markers indicate the evolution of the state in
	pressure-volume space across the ZND wave of PBX9502. The green and black
	markers represent the state of the reactants and products respectively. The
	red line is the pressure reference curve for the product EoS and represents
	the principal isentrope. The red dashed line represents the Rayleigh line.}
	\label{fig:znd_pbx9502}
\end{figure}

The ZND model for the structure of one dimensional detonation waves can be used
to calculate the Rayleigh line, the von Neumann spike, the CJ state and the
principal isentrope of the rarefaction wave. This can be done using solely the
EoS models of the reactants and products, and the Rankine-Hugoniot
conditions.
Figure \ref{fig:znd_pbx9502} shows the ZND wave for PBX9502
and compares the numerical simulation of a one-dimensional detonation wave with
the calculated Rayleigh line and the principal isentrope from the ideal
detonation code.

The blue markers indicate the evolution of the state in the explosive as a
whole. The ambient depressurised state is in the bottom right. Across the shock
wave (approximately three cells) the state approaches the predicted von
Neumann spike (top left). As the explosive burns, the state of the explosive
(blue) follows the Rayleigh line towards the CJ state. During this stage, the
explosive is a mixture of reactants and products, which are at pressure
equilibrium but have different specific volumes.
The green and black markers represent the states of each material.
Following the CJ state, the explosive consists entirely of products and rarefies
following the principal isentrope.

The blue markers lie on the Rayleigh line as expected, and the rarefaction of
the detonation products follows the principal isentrope as used for the
calibration. This demonstrates that the calibration process is working as
expected. Furthermore the speed of propagation of the wave is observed to be as
predicted by the Rankine-Hugoniot equations.

Note that even
in the reaction zone the state of the products lies roughly on the
principal isentrope. This indicates that there is not much heat transfer
between reactants and products in the reaction zone. The isentropic closure law
could therefore be applied in place of temperature equilibrium leading to
similar results.

\begin{figure}
	\centering
	\includegraphics{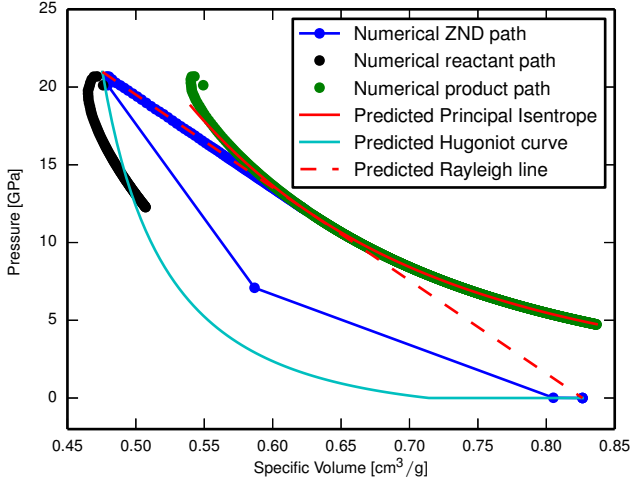}
	\caption{The one-dimensional ZND wave for EM120D is presented as for Figure
	\ref{fig:znd_pbx9502}. Note that the Hugoniot curve is flat for specific
	volumes larger than the crushing volume of 0.71 $\unit{cm^3g^{-1}}$ as a
	result of the porosity model.}
	\label{fig:znd_em120d_1000}
\end{figure}

However, this is not true in general. Figure \ref{fig:znd_em120d_1000} shows the
equivalent plot for the emulsion EM120D. The product density at
the front of the reaction zone places the state above the principal isentrope.
This shows that in the first stage of the burning the temperature equilibrium
constraint causes heat to transfer from products to reactants, and the
reactants compress to a higher density than that corresponding to the von
Neumann spike.

\begin{figure}
	\centering
	\includegraphics{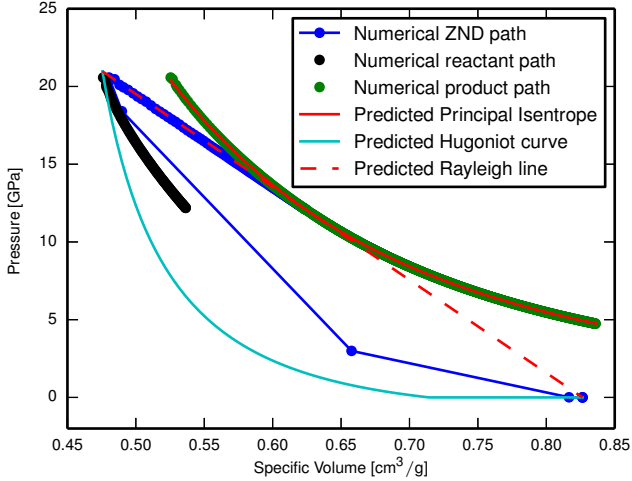}
	\caption{The one-dimensional ZND wave for EM120D is presented once again,
	but this time with an increased heat capacity for the reactants. This has
	the effect of lowering the von Neumann spike temperature.}
	\label{fig:znd_em120d_1500}
\end{figure}

The extent to which heat transfer between reactants and products occurs is
dependent on the temperature at the von Neumann spike as
predicted by the EoS of the reactants, and how this temperature compares to the
temperatures on the principal isentrope of the product EoS. To illustrate this
dependence, Figure
\ref{fig:znd_em120d_1500} shows the ZND wave for the emulsion EM120D but
with the specific heat capacity of the reactants arbitrarily increased from
$1000\ \unit{JK^{-1}kg^{-1}}$ to $1500\ \unit{JK^{-1}kg^{-1}}$.
This change reduces the von Neumann spike temperature in the reactants and
alters the behavior to be analogous to what is observed in the ideal explosive
PBX9502 in Figure \ref{fig:znd_pbx9502}.
Depending on the heat capacities and other parameters which
are known with little precision, the difference between a thermal equilibrium
condition and an isentropic closure law may be of little significance. It is
thus not possible to conclude definitively whether the thermal equilibrium
condition is physically justified or otherwise.

\section{Predictive Modeling of Detonation Waves in Rate Sticks and Slabs}
\label{sec:pred_mod}

In one-dimensional domains the velocity of the detonation wave is the CJ
velocity, which depends solely on the EoS of the products and the ambient
density of the explosive. However the VoD measured in rate
sticks and slabs is significantly reduced from the ideal VoD,
as a result of the loss of energy to the confining material. The measured
VoD therefore depends on the geometry and material of the
confiner and will also depend on the reaction rate. Near-ideal
explosives, with a very fast reaction rate, deviate from the ideal VoD to a
lesser degree than non-ideal explosives with slower reaction rates.

It is currently impractical to model the chemistry of the reaction directly.
A one-step reaction model is used to calculate the rate at which reactants
transition to products. The parameters for this model must be calibrated using
experimental VoD measurements. This does not preclude the
ability to be predictive. It was found that in practice only a few measurements
are required for the calibration of the parameters, and that these same
parameters can be used to predict detonation velocities in a different
context.

For the ideal explosive, PBX9502, the reaction rate is calibrated using
VoD data for rate sticks of multiple radii. The resulting
parameters are used to predict the VoD in slabs of varying
thickness. For the less ideal explosive EM120D, data
for unconfined rate sticks (air confinement) are used for the calibration. Predictions are
then made for rate sticks confined by concrete and steel.

The simulation of rate sticks is carried out in two dimensions using the
assumption of rotational symmetry about the axis of the rate stick. This
requires the use of a geometric source term \cite{Michael2015a}.

The detonation wave is initiated using a booster - an area of high pressure gas
which shocks the explosive, initiating the reaction. After the start of the
simulation, the detonation wave must be modeled for some time to allow it to
settle to its steady speed. After the wave has converged, the speed can be
measured by simply observing the distance covered in some time interval. The
measurement of the position of the shock wave introduces an error related to
the discretisation of the grid. However the error in the speed measurement can
be reduced by measuring the speed over longer time intervals.

For unconfined rate sticks and slabs, the confining air is modeled with a polytropic EoS
with an adiabatic gamma of $1.4$. For rate sticks of EM120D with solid
confinement, the shock \MG EoS is used with the same parameters as Schoch \etal
\cite{Schoch2013}. Note that while the mathematical
formulation uses a temperature equilibrium condition between reactants
and products, only pressure equilibrium is used between the explosive and
confiner \cite{Michael2015a}. As such, temperatures in the confiner are inconsequential and use of
the shock \MG EoS is appropriate.

A linear fit of experimental data for the shock speed provides the parameters
$a$ and $b$ \eqref{eq:linear_Du}.
The reference curves are
\begin{align}
	p_{\REF} &= \frac{\rho_0 \chi a^2}{(1-b\chi)^2} \label{eq:pref_bare} \\
	e_{\REF} &= \frac{1}{2} (v_0 - v) p_{\REF} \\
	\rho \Gamma &= \rho_0 \Gamma_0, \label{eq:rhoGamma_smg}
\end{align}
where $\Gamma_0$ is the ambient \Gruneisen gamma, $\rho_0 = 1/v_0$ is the
initial density and $\chi$ is defined as
\begin{equation}
	\chi = 1 - \frac{\rho_0}{\rho}.
\end{equation}
The parameters are given in Table \ref{tab:smg_all}.
\begin{table}
	\centering
	\begin{tabular}{|r|cccc|}
		\hline
		& $\rho_0 [\unit{kgm^{-3}}]$ & $a [\unit{ms^{-1}}]$ & $b$
			& $\Gamma_0$ \\
		\hline
		Steel \cite{Thiel1977} & 7840 & 3670 & 1.645 & 2.0 \\
		Concrete \cite{Chhabildas1995} & 2340 & 2235 & 1.745 & 2.0 \\
		\hline
	\end{tabular}
	\caption{Parameters for the shock \MG EoS for modeling confinement.}
	\label{tab:smg_all}
\end{table}


\subsection{Calibration of Reaction Rate for EM120D}
\label{sec:calib_em120d}
The reaction rate is very difficult to measure experimentally or
to evaluate on the basis of chemical arguments. In reality the explosive does
not transition directly from reactants to products but undergoes many
intermediate reactions associated with varying amounts of energy. For the
purposes of the simulation, these processes are combined into a single
pressure-dependent expression, $K$, for the reaction rate as is used by Schoch
\etal \cite{Schoch2013},
\begin{multline}
	K = \frac{\diff \lambda}{\diff t} = -\lambda^{indx} \bigg(
	p \frac{1-a_h}{\tau_s} \\ + H(p-p_h)
	\frac{a_h}{\tau_h} \left(\frac{p[\unit{Pa}]}{10^9}\right)^{N_p} \times 10^9
		\bigg),
	\label{eq:reaction}
\end{multline}
where
\begin{equation*}
	a_h = \exp \left( - \left(\frac{1-\lambda}{\omega_h} \right)^{N_a} \right).
\end{equation*}

The leading coefficient causes the reaction rate to slow as the reaction nears
completion. The regression index of the reaction is $indx$. The second term
represents the hotspot reaction, where $H(x)$ is a Heaviside function, and
$p_h$ is the critical pressure required for ignition. The first term is a bulk
burning term which determines the reaction rate once the explosive is fully
ignited. The parameter $a_h$ is initially 1, causing the hotspot reaction term
to dominate. As the reaction progresses, $a_h$ approaches zero, and the
equation becomes dominated by the bulk burning term.

$\omega_h$ determines the degree to which the hotspot process consumes the
available explosive. $\tau_s$ and $\tau_h$ determine the time scales
of the reaction, and the constant $N_a$ controls the speed at which the
hotspots transition to a bulk burning process. Note that it is the pressure in
$\unit{GPa}$ which is raised to the power of $N_p$.

The calibration was carried out using data from Dremin
\cite{Dremin1999} (which is also used by Schoch \etal \cite{Schoch2013}) for the
weakly confined rate sticks. It is then demonstrated that the
same parameters allow for predictions to be made for confined detonation
waves. The only input required for the predictions is the EoS of
the confining material.

It was found that the principal parameters affecting the VoD
were $\tau_h$ and $N_p$. The other
parameters were assigned the same values as were used by Schoch \etal
\cite{Schoch2013}. Since there were only two degrees of freedom in the
calibration process, only two data points were required to fully constrain the
system. These were chosen to be the detonation velocities for $20\ \unit{mm}$ and
$30\ \unit{mm}$ rate sticks which were $4920\ \unit{ms^{-1}}$ and $5470\
\unit{ms^{-1}}$ respectively \cite{Dremin1999}.

A two-dimensional implementation of the secant method was applied to minimize
the discrepancy between the numerical results and the
experimental data. For each radius three evaluations of the velocity with
different parameters are required to construct a two-dimensional plane in three
dimensional space relating the values of the parameters with the VoD. The
intersection of this plane with the experimental VoD constitutes a line through
the two-dimensional parameter space. The
final step is to find the intersection between this line and a similarly
calculated line for the second value of the radius.

This process is repeated iteratively until good agreement with the experimental
velocities is found. The results of the calibration along with the other
parameters are presented in Table \ref{tab:reaction_rate_params}.
\begin{table}[h]
	\centering
	\begin{tabular}{|c|c|}
		\hline
		$\tau_h$ & $13\ \unit{\mu sGPa}$ \\
		$\tau_s$ & $20\ \unit{\mu sGPa}$ \\
		$p_h$ & $1.51\ \unit{GPa}$ \\
		$indx$ & 0.667 \\
		$\omega_h$ & 0.95 \\
		$N_a$ & 9.0 \\
		$N_p$ & 1.11 \\
		\hline
	\end{tabular}
	\caption{Parameters for the reaction rate model for EM120D. See equation
		\eqref{eq:reaction}.}
	\label{tab:reaction_rate_params}
\end{table}

\begin{figure}
	\centering
	\includegraphics{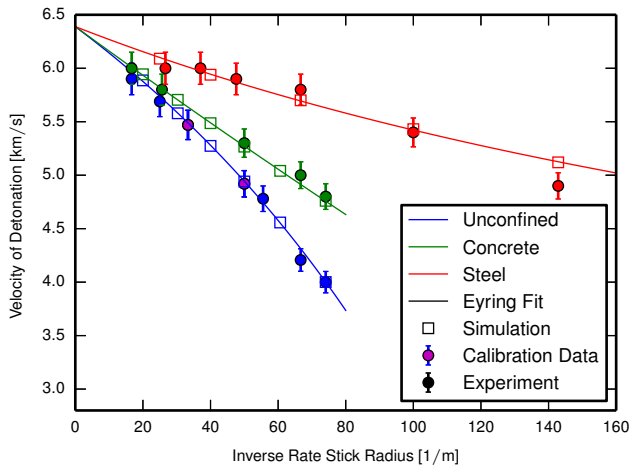}
	\caption{The plot shows the radial dependence of the VoD
	for rate sticks of EM120D. The lines interpolate the numerical results
	(square markers) using Eyring fits. The circular markers with error bars
	are the experimental data \cite{Dremin1999}. The experimental data for
	unconfined rate sticks that were used for the calibration are highlighted
	with magenta markers.
	}
	\label{fig:vod}
\end{figure}

Figure \ref{fig:vod} shows the simulation results as squares for the
calibrated reaction rate. These results are fit using the Eyring equation
\cite{Campbell1976}
\begin{equation}
	D = D_{CJ} \left(1 - \frac{A}{R-R_C} \right).
	\label{eq:eyring}
\end{equation}
The resulting values for $A$ and $R_C$ are in Table \ref{tab:fits}.

\begin{table}
	\centering
	\begin{tabular}{|c|cc|}
		\hline
		Confiner & $A [\unit{mm}]$ & $R_C [\unit{mm}]$ \\
		\hline
		Air      & $3.73$ & $3.5  $ \\
		Concrete & $3.61$ & $-0.6 $ \\
		Steel    & $1.94$ & $-2.8 $ \\
		\hline
	\end{tabular}
	\caption{Parameters for the fits of the radial dependence of the VoD with
	rate stick radius for EM120D, with ideal VoD $D_{CJ} = 6.3895\
	\unit{kms^{-1}}$.}
	\label{tab:fits}
\end{table}

Figure \ref{fig:vod} shows good agreement between the predictions from the
Eyring fits and the experimental data \cite{Dremin1999}. The lines
for steel and concrete show good agreement over a wide range of radii with the
exception of the narrowest steel confined rate stick. This is consistent with
the results of Schoch \etal \cite{Schoch2013}. Furthermore the
detonation velocities for unconfined rate sticks show good agreement for all radii
despite the calibration having been done using the data points for
$20\ \unit{mm}$ and $30\ \unit{mm}$ radius only. The Eyring fits work well for the
simulation data, which means that the simulations are converging towards the
ideal VoD in the large radius limit as is expected.

The fact that the calibration was successful using just two parameters and two
data points demonstrates that the physics of the detonation waves is being captured
well by the EoS models and the MiNi16 formulation. It also suggests
that it may be possible to use an expression much simpler than \eqref{eq:reaction}
for the reaction rate and achieve the same predictive capability.

\subsection{Calibration of Reaction Rate for PBX9502}

For PBX9502 we use a simplified version of the ignition and growth model
presented by Tarver and McGuire \cite{Tarver2002} and used by Wescott \etal
\cite{Wescott2005}. We calibrate the reaction rate model using VoD data for
unconfined rate sticks \cite{Jackson2015}. The model is used to predict the VoD
in slabs of varying thickness. The predictions are then compared with
experimental data \cite{Jackson2015}.

The form of the reaction rate was chosen to be
\begin{align}
	K &= r_{DG} S_{G}(\lambda),
\end{align}
where
\begin{align}
	r_{DG} &= k_{DG} (1-\lambda)^{1/3} \lambda^{N_\lambda} \\
	S_{G} &= \frac{1}{2}(1 - \tanh(30(0.1-\lambda))).
\end{align}
However we believe that the results
presented here are compatible with a pressure dependent model. This could
be achieved through modification of the form of the reaction rate or adjustment
of the exponents.

The calibration was carried out using the same methodology as was applied for
the emulsion in the section \ref{sec:calib_em120d}. In this case the two free parameters
are $k_{DG}$ and $N_\lambda$. The
final value for $k_{DG}$ was $60.65\ \unit{\mu s^{-1}}$, while $N_\lambda$ was
$1.56$.

\begin{figure}
	\centering
	\includegraphics{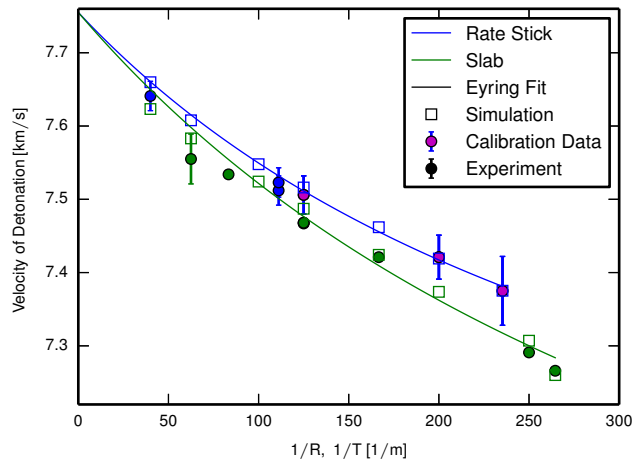}
	\caption{The dependence of the VoD with size is presented for both rate
	sticks and slabs. The lines are Eyring fits through the simulation data
	(square markers), while the markers represent the experimental data
	\cite{Jackson2015}. Note that for slabs the $x$ axis represents the inverse
	thickness, where the thickness is measured across the whole slab, while for
	rate sticks the radius, not the diameter, is used.}
	\label{fig:vod_pbx}
\end{figure}

The results are shown in Figure \ref{fig:vod_pbx}.
The parameters for the Eyring fits are given in Table \ref{tab:fits_pbx9502}.

\begin{table}
	\centering
	\begin{tabular}{|c|cc|}
		\hline
		Confiner & $A [\unit{mm}]$ & $R_C [\unit{mm}]$ \\
		\hline
		Rate sticks & $0.34$ & $-2.8$ \\
		Slabs & $0.37$ & $-2.3$ \\
		\hline
	\end{tabular}
	\caption{Parameters for the fits of the radial dependence or thickness
	dependence of the VoD of PBX9502, with $D_{CJ} = 7.755\ \unit{kms^{-1}}$.}
	\label{tab:fits_pbx9502}
\end{table}

\section{Conclusions}
\label{sec:conc}
The objective of this work is to improve the robustness and accuracy of
simulations of ideal and non-ideal explosives by introducing
temperature dependence in mechanical EoS models for the reactants and products.


\begin{figure}
	\centering
	\includegraphics{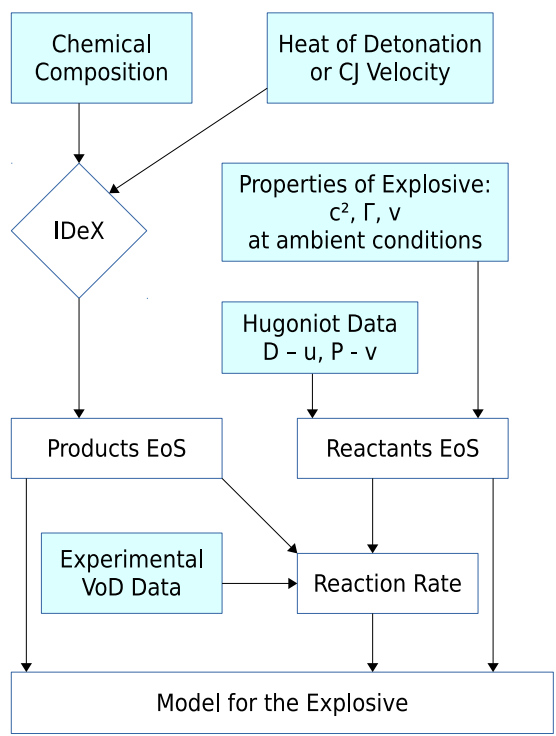}
	\caption{The steps in the methodology for building a model for an explosive
	are summarized here. The elements in colored boxes represent
	data which must be acquired experimentally. The box labeled \emph{IDeX}
	represents an ideal detonation code. The resulting model can be
	used quantitatively to calculate detonation speeds for explosives in new geometries, with
	new confining materials or of different dimensions.}
	\label{fig:method}
\end{figure}

A methodology for constructing a model for generic explosives has been presented. This is
summarized in Figure \ref{fig:method}, which lays out the experimental data
required and outlines the steps involved in the methodology's application.

The reactant EoS was developed following Davis \cite{Davis2000} and is
calibrated using experimental data for the Hugoniot curve, thus reproducing the
desired shock-response behavior. The temperatures are derived solely from the
Hugoniot data and the thermodynamics of the explosive in ambient conditions,
since there is very limited thermal data available for explosive reactants.
The EoS explicitly accounts for the influence of porosity on the
post-shock temperatures using the snow plow model.

The product EoS is an adaptation of the JWL EoS which accommodates
evaluation of the temperature.
The reference curves are calibrated to data for the principal isentrope
from the ideal detonation code \emph{IDeX} \cite{Braithwaite1996}.
The ideal detonation code requires
the chemical composition of the explosive as well as the energy content of the
explosive in comparison to the detonation products. Note that if the energy
content is unknown, then an experimental measurement of the ideal VoD can be
used instead.

Use of the ideal detonation code not only permits the calculation of
temperatures but more accurate values for a volume-dependent \Gruneisen
gamma. This is important, since in non-ideal detonation waves the state of the
products is expected to lie below the reference curve of the EoS.
Away from the reference curve, the validity of the EoS relies on an accurate
expression for the \Gruneisen gamma.



The methodology was applied to the non-ideal explosive emulsion
EM120D and the ideal TATB based explosive PBX9502.
The resulting models for the explosives were used in the context of the MiNi16
formulation \cite{Michael2015a} to perform direct numerical simulation of the
detonation wave and its interaction with the confiner.
Results demonstrate that the solution of the nonlinear temperature equilibrium
equation can be found robustly.

For EM120D, the predictive capability demonstrated by Schoch \etal
\cite{Schoch2013} was successfully reproduced. The methodology was shown to be
capable of predicting the effect of strong confinement on the VoD, despite
using only data for weakly confined rate sticks in the calibration
process.
For PBX9502, the methodology was applied to predict the dependence of the VoD on
the geometry. The model was calibrated using rate stick data, and used to
predict the VoD in a slab geometry. In each case, the predictions were verified
using experimental data.

\begin{acknowledgments}
S.D.\@Wilkinson acknowledges financial support from the EPSRC Centre for
Doctoral Training in Computational Methods for Materials Science under grant
EP/L015552/1. The authors are grateful to Alan Minchinton for useful
discussions and provision of data generated using \emph{IDeX}.
\end{acknowledgments}

\bibliography{library}

\end{document}